\documentclass[conference]{IEEEtran}
\IEEEoverridecommandlockouts
\usepackage{cite}
\usepackage{amsmath,amssymb,amsfonts}
\usepackage{algorithmic}
\usepackage{graphicx}
\usepackage{textcomp}
\usepackage{xcolor}
\usepackage[linesnumbered,ruled,noend]{algorithm2e}
\usepackage{multirow,multicol}
\usepackage{comment}

\AtBeginDocument{%
	}
				
\newcommand{\Paragraph}[1]{~\vspace*{-0.9\baselineskip}\\{\bf #1}}
\usepackage{url}

\begin{document}

\title{WindVE: Collaborative CPU-NPU Vector Embedding}

\begin{comment}
\author{\IEEEauthorblockN{Anonymous Authors}}
\end{comment}

\author{
\IEEEauthorblockN{Jinqi Huang*}
\IEEEauthorblockA{\textit{Huawei Technologies Co., Ltd} \\
huangjinqi1@huawei.com}
\and
\IEEEauthorblockN{Xuebing Yu*\thanks{*These authors contributed equally to this work.}}
\IEEEauthorblockA{\textit{Huawei Technologies Co., Ltd} \\
yuxuebing3@huawei.com}
\and
\IEEEauthorblockN{Li Zeng}
\IEEEauthorblockA{\textit{Huawei Technologies Co., Ltd} \\
zengli43@huawei.com}
\and
\IEEEauthorblockN{Entong Li}
\IEEEauthorblockA{\textit{Huawei Technologies Co., Ltd} \\
lientong@huawei.com}
\and
\IEEEauthorblockN{Zhixiong Ning}
\IEEEauthorblockA{\textit{Huawei Technologies Co., Ltd} \\
ningzhixiong1@huawei.com}
\and
\IEEEauthorblockN{Jinhua Zhou}
\IEEEauthorblockA{\textit{Huawei Technologies Co., Ltd} \\
zhoujinhua1@huawei.com}
\and
\IEEEauthorblockN{Rongqian Zhao}
\IEEEauthorblockA{\textit{Huawei Technologies Co., Ltd} \\
zhaorongqian@huawei.com}
\and
\IEEEauthorblockN{Xin Chen}
\IEEEauthorblockA{\textit{Huawei Technologies Co., Ltd} \\
chenxin@huawei.com}
}

\maketitle

\thispagestyle{plain}
\pagestyle{plain}

\begin{abstract}
Retrieval-Augmented Generation (RAG) enhances large language models (LLMs) using information retrieval and vector embeddings. This improves context for text generation but can impact cost-performance, as vector operations consume significant latency (up to 20\%). Optimizing computational resources for vector embeddings is thus crucial for cost-effective inference.

This paper analyzes vector embedding deployment costs and finds that boosting concurrent query processing is key to reducing them. To achieve this without sacrificing performance, we designed a queue manager to offload CPU peak queries using linear regression for optimal queue depths. We also developed WindVE, a CPU-NPU system leveraging processor differences to manage traffic surges. Experiments show WindVE achieves 22.3\% higher concurrency than a non-offloading scheme, translating to 1.22x throughput under the same hardware or up to 18.6\% hardware cost savings for the same performance.
\end{abstract}


\section{Introduction}\label{sec:introduction}
\begin{figure*}[!ht]
	\centerline{\includegraphics[width=0.8\textwidth]{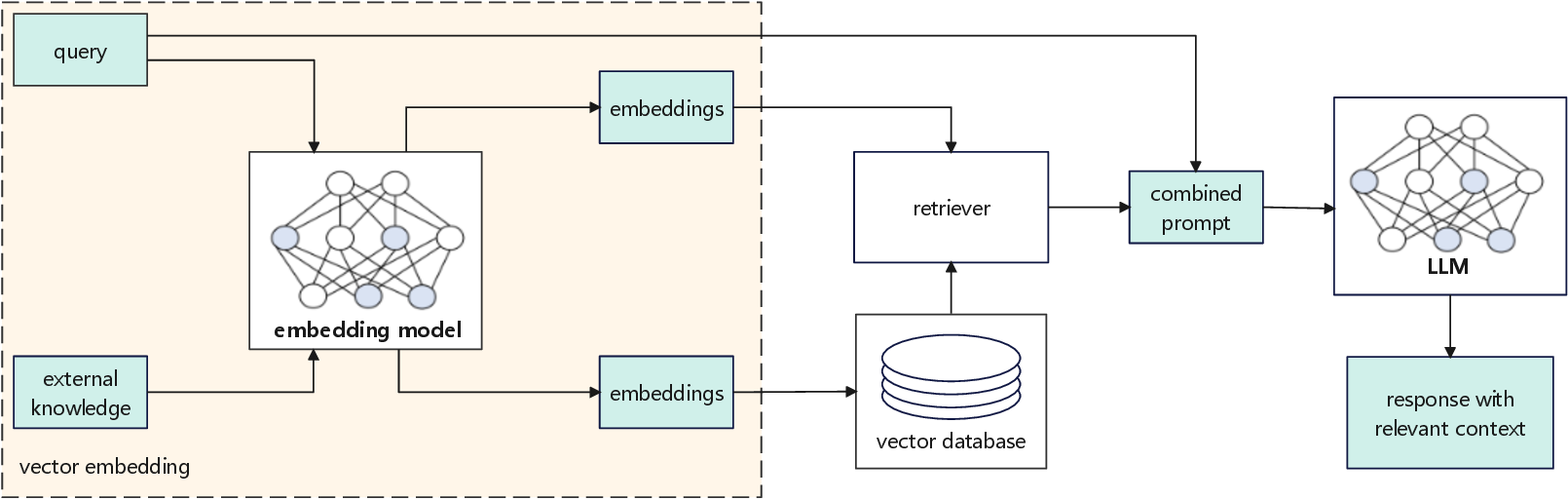}}
	\caption{RAG workflow.}
	\label{fig:rag}
\end{figure*}

In today’s landscape, LLMs are increasingly gaining prominence and are extensively utilized across a multitude of industrial applications, including the Huawei O3 Knowledge Community \cite{site:O3} and OpenAI’s ChatGPT \cite{site:chatgpt}. RAG is an innovative approach in LLM applications that enhances the accuracy \cite{LLM-RAG} and prevents hallucination \cite{hallucination} of LLMs by integrating information retrieval systems \cite{BGM,CoQ,PaperQA,IAG,NoMIRACL}. 
As a core component of RAG, vector embedding uses numerical representations that convert text into high-dimensional vectors, capturing semantic meanings and relationships between words or sentences \cite{context-embeddings,Chat2Data}. 
In the context of RAG, vector embedding plays a vital role in both retrieval \cite{VectorSearch} and generation \cite{ve-generation} phases of LLMs. 
Figure \ref{fig:rag} depicts a typical RAG workflow. 
During retrieval, an embedding model extracts relevant information from an external database. These embeddings are then fused with the query to guide the LLM, improving the generation’s context and specificity. Embeddings are also used in other RAG modules like query rewriting. Due to this online, per-request embedding requirement, vector embedding services face massive usage (millions of calls per month), unlike traditional offline systems.

Nevertheless, from the business point of view, LLM inference services are highly sensitive to cost–performance ratio. 
The high cost associated with inference system deployment can lead to a high price for customers, ultimately resulting in a loss of product competitiveness. 
For example, the price of the lately released LLM inference service gpt-4o-mini is as low as \$0.150 / 1M input tokens \cite{o1-mini-price}. 
Specifically, in modern industry applications, vector embedding and retrieval consume up to 20\% of the total latency within a typical RAG-integrated LLM inference system \cite{rag-survey}. 
Therefore, companies must strive to optimize vector embeddings to ensure that they offer the best possible performance at a reasonable price. %In this paper, we have found out that improving the system capacity to process concurrent queries while ensuring the SLOs are met is the key to reducing service costs and enhancing the product's cost-effectiveness. The deduction of this conclusion will be explained explicitly in a following section.  

In this paper, We conduct an analysis of the cost composition of vector embeddings and found that improving the max amount of concurrent queries that an inference system can process is crucial to enhancing the cost–performance ratio of the product. By focusing on boosting the capability to process concurrent peak queries, the product's cost–performance ratio can be significantly optimized. Therefore, we first design a queue manager to facilitate seamless CPU-NPU collaboration for vector embedding. Furthermore, we offer a linear regression based estimator to guide the calibration of its pivotal parameter, aiming to maximize concurrency while meticulously preserving latency performance. We further design WindVE, a system to offload peak concurrent requests using a CPU-NPU heterogeneous architecture. Note that this approach can also be applied to CPU-GPU heterogeneous architecture or other hardware platforms with AI processors \emph{and} CPUs from all manufacturers, though we use `CPU-NPU collaboration' in this paper for simplicity. WindVE leverages the strengths of both CPU and NPU to efficiently handle surge in traffic, ensuring optimal throughput and cost efficiency. Finally, we conduct experiments based on the proposed scheme and achieved the concurrency and the throughput of up to 22.3\%$\times$ higher than the non-offloading scheme. The experimental results indicate that the proposed scheme can achieve a maximum cost-effectiveness of 1.22$\times$ and save 18.6\% of deployment costs. Our contributions can be concluded below:
\begin{itemize}	
\item We designed a queue manager to facilitate seamless CPU-NPU collaboration for vector embedding, coupled with a linear regression based estimator for ascertaining its critical parameters to maximize concurrency.	
\item We developed WindVE, a flexible and scalable system framework for high-throughput offloading vector embedding with CPU-NPU collaboration.
\item We demonstrated WindVE incorporating CPU-NPU heterogeneous computing for peak query offloading in vector embedding applications. Our experimental results demonstrate that we have achieved 1.20$\times$ concurrency with service-level objectives (SLOs) met.
\end{itemize}
This paper is organized as follows: Section \ref{sec:background} reviews hardware efficiency for vector embeddings. Section \ref{sec:Motivation} analyzes why system concurrency reduces costs and explores CPU offloading benefits. Section \ref{sec:algorithm} details our WindVE framework for peak query offloading. Section \ref{sec:experiment} demonstrates WindVE’s improvements in concurrency and throughput. Section \ref{sec:conclusion} concludes.

\section{Background}\label{sec:background}
In this section, the historical work are presented, including vector embedding performance improvement, model inference acceleration and CPU-NPU/GPU collaboration.

\subsection{Vector Embedding}

A group of literature related to vector embedding aim to improve functionality: Ref \cite{image-text-retrieval} proposed an embedding framework using a memory network to obtain cross-modal memory-enhanced embeddings; Ref \cite{fastrp} introduces a scalable algorithm for network embeddings to capture transitive relationships in a graph; Ref \cite{discriminative-embedding} proposed an embedding method extending random projection to an average language vector to improve document embeddings discriminating ability; Ref \cite{Sequencing-embedding} proposed an alignment-free embedding approach to generate a fixed-length feature vector representation from the raw sequencing reads without assembly; Ref \cite{adaembedded} proposed adaptive embedding method to distinguish important features in dynamic data distribution in training; and Ref \cite{freddy} exploited how to utilize word embeddings to augment and enrich queries in database management systems. 

Another group of designs aim to improve the efficiency of vector embedding system: Ref \cite{nonlinear-embedding} employed nonlinear embeddings to project data points in computer vision area to low-dimensional space and proved to be more efficient than other embedding algorithms in finding the exact nearest neighbor based on the Euclidean distance; Ref \cite{multitask-hashing, binary-hashing, compositional-embedding} used hashing with collisions to compress embeddings to achieve higher processing efficiency; and Ref \cite{cafe} accelerated embeddings entailing only hash processes with one additional potential embedding lookup. 

These work all improve functionality or performance by modifying embedding algorithms, which may have unpredictable impact on the effect of business applications.
Our work focuses on lossless acceleration of vector embedding, thus do not consider these algorithms as counterparts.

\subsection{Inference Acceleration}
A large group of works has contributed to LLM inference acceleration, such as ClossalAI \cite{ColossalAI}, vLLM \cite{vLLM}, Triton \cite{triton}, etc. ClossalAI provides an inference framework built upon Colossal-AI, to support parallelism for large-scale models, pre-built large models, engine encapsulation, and an online service system; vLLM gives an efficient and easy-to-use LLM inference and serving library with cutting-edge inference acceleration technologies such as paged attention \cite{pagedattention}, continuous batching, speculative decoding \cite{speculative}, etc; Triton is an inference server introduced by NVIDIA for standardized AI model deployment and execution empowered by dynamic batching, concurrent execution, optimal configuration, and streaming audio and video. 
These frameworks all implement inference acceleration using LLM characteristics, such as billions of parameters, parallelism among multiple cards, KVCache mechanism and prefill/decoding phases.
In contrast, the embedding inference usually adopts a model smaller than 1B, without the need of KVCache and multi-card parallelism, as it can be well placed and processed by a single card or CPU cores.
Also, the output of an embedding inference is a tensor instead of a word sequence, thus the decoding phase does not exist in this scenario. 
Therefore, existing LLM inference methods are not considered as major counterparts, as they cannot reproduce the same level of performance boost in embedding inference.   

For embedding model acceleration, FlagEmbedding \cite{site:flagembed} is the state-of-the-art solution for vector embedding. Serving as an open-source library for enhancing sentiment representations of words or sentences, FlagEmbedding uses vector embedding to capture the sentiment relationships between tokens for more precise text understanding and information retrieval. However, it has some major drawbacks: it does not efficiently reduce high dimensions, causing sparse representations in high-dimensional space; it adds embedding in training phase, leading to high computational costs; it does not support processing in CPUs and NPUs at the same time. The first two shortcomings are introduced by FlagEmbedding's embedding algorithm, whose improvement is not in scope of this paper, while overcoming the last fits the purpose of this paper, thus selected as the counterpart in our experiments in Section \ref{sec:experiment}.

\subsection{CPU-NPU Collaboration}

CPU-NPU heterogeneous computing has been widely used in applications such as power systems \cite{power-system}, pollution dispersion \cite{polution}, queue processing system \cite{queue-processing}, image processing \cite{image-processing}, speech processing \cite{speech-processing}, federated edge learning \cite{federated}, and pattern matching \cite{pattern-matching}. 
The detailed techniques of collaboration include coarse-grain thread-level parallelism across CPUs and NPUs/GPUs \cite{parallel-proc}, CPU-NPU workload partitioning \cite{node-embedding, federated}, and heterogeneous tensor or pipeline parallelism \cite{hetegen}. 

These work all aim to reduce the responding latency of model inference, rather than the maximum concurrency, which is the focus of this paper.
These two objectives are nearly contradictory, as increasing the batch size often leads to higher latency. 
Besides, the number of parameters in embedding models is usually lower than 1B (e.g., BGE 326M \cite{bge}, Jina 570M \cite{jina}).
Thus, vector embedding can be run as an entire instance on CPU, eliminating the need for complex model slicing like Ref \cite{hetegen}.

\section{Motivation}\label{sec:Motivation}

In this section, we first illustrate the motivation for this work by analyzing why the system’s capacity to process the maximum number of concurrent queries is pivotal in reducing the expenses associated with inference services. This analysis underpins our proposal to leverage CPUs that are already present in GPU/NPU server configurations to offload peak query loads, thereby enhancing concurrency. This strategy can significantly decrease the cost of request processing.

\subsection{Why Concurrency Matters} \label{sec:concurrent}
The latency of an inference service is defined as the total time taken from receiving a query to producing the corresponding output. The maximum concurrency of an inference service is the maximum number of concurrent queries the system can process given the SLOs and memory constraints. The throughput is the number of queries processed within a time window.
In industry applications, latency is a key performance indicator related to SLOs and user experience; maximum concurrency indicates the system’s instantaneous capacity to process query bursts; and throughput defines the average performance of query processing within the given time window, which can be affected by the actual incoming query distributions.
\begin{figure}[h!]
	\centerline{\includegraphics[width=0.9\linewidth]{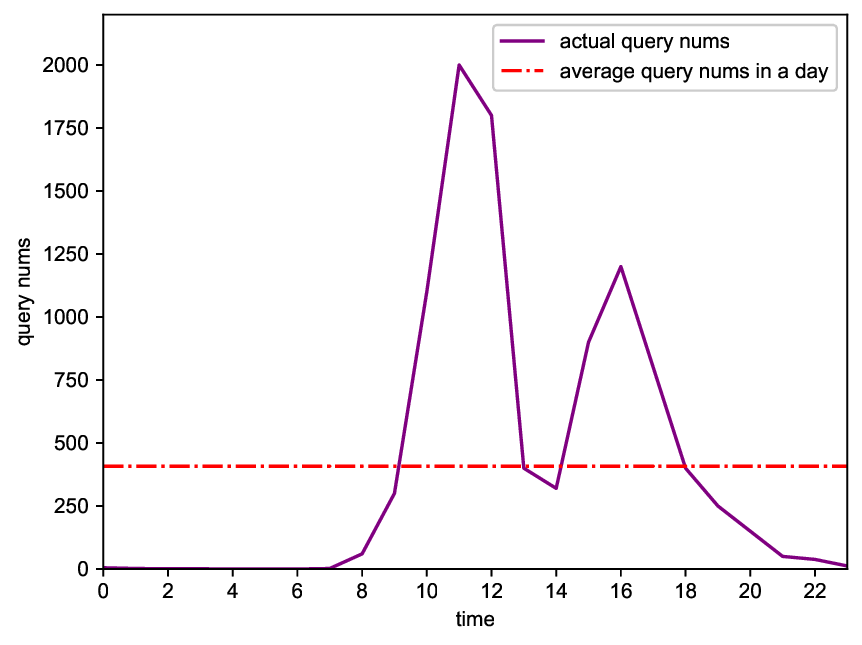}}
	\vspace{-0.4cm}
	\caption{An illustration of query number changes in a day.}
	\label{fig:workload}
\end{figure}

There are two ways of deploying an inference service. 
One is to deploy the service according to the received query number within the time window and the throughput. 
To express service deployment costs using this approach, the maximum acceptable latency and the actual total latency are denoted as $t_{total}^{max}$ and $t_{total}$, respectively. 
We represent the $t_{total}^{max}$ and $t_{total}$ as the following expressions:
\begin{gather}
	t_{total}^{max} = t_{wait}^{max} + t_{proc}    \\
	t_{total} = t_{wait} + t_{proc} 
\end{gather}
Where $t_{proc}$ represents the average processing time in an inference system, and $t_{wait}^{max}$ and $t_{wait}$ are the maximum acceptable waiting time and the actual waiting time when taking average time to process a query. 
To avoid timeout, the following condition has to be met:
\begin{equation}
	t_{wait} \leq t_{wait}^{max} = t_{total}^{max} - t_{proc}
\end{equation}
Here, we define $n$ as the number of other queries processed when a query is waiting in the queue:
\begin{equation}
	n = \left \lfloor \frac{t_{total}^{max} - t_{proc}}{t_{proc}} \right \rfloor 
\end{equation}
Supposed the query number received in a second for an application is $N$, the deployment cost of an inference service can be expressed as follows: 
\begin{equation}\label{eq:cost}
	\text{Cost} = \frac{\frac{N}{n}}{T} \cdot D \cdot P
\end{equation}
Where $D$ and $P$ are device number per instance and price per device, respectively. 
Service deployment according to the query number with a time window using Equation \ref{eq:cost} is cost-saving because it utilizes hardware resources thoroughly. 
However, in typical industrial applications,  online query numbers varies significantly during a day (as shown in Figure \ref{fig:workload}). 
Using an average processing ability $T$ to estimate the number of processors to deploy does not consider potential query bursts that excess the service capacity, which lead to overshoot in processing time. 
In the worst case, query bursts with number above average have risk of not meeting SLOs due to large waiting time. 

In contrast, another way of deploying an service is to use the peak query number and the system capacity of processing the maximum amount of queries. 
The following equation gives the deployment cost using this approach:
\begin{equation}\label{eq:cost1}
	\text{Cost} = \frac{N_{peak}}{C} \cdot D \cdot P
\end{equation}
Where $N_{peak}$ and $C$ represent the peak query number and the system maximum concurrency. 
Using Equation \ref{eq:cost1} to deploy a service can avoid overtime processing, but hardware resources are not fully utilized if the query number is smaller than $N_{peak}$. 

All in all, using average processing ability to deploy can cover average scenarios, whilst using maximum concurrency ensures performance quality when peak query bursts occur. 
These two ways both have many applications in industry:
\begin{itemize}
\item average concurrency is suitable for  delay-insensitive scenarios such as customer service and query answering, where requests can be waiting in queue and the latency of each single request is not guaranteed;
\item maximum concurrency is widely used in serious business scenarios, such as fault maintenance and many popular Internet applications (e.g., Tiktok and rednote), which requires that the latency of each request is strictly not higher than the given border, also knows as SLOs.
\end{itemize}  
An ideal solution would be using the average concurrency to deploy whilst make of the ready-exist hardware resources to extend the maximum concurrency without extra costs. 
Meanwhile, most AI accelerators, such as NPUs and GPUs, are assembled with multi-core CPUs in servers. 
From practical observation, these CPUs only run basic service framework without operating computing-intensive tasks, causing low utilization rate of less than 10\%. 
Hence an scheme of using CPUs to offload peak queries with NPUs processing average queries is proposed. 
This scheme can fully utilize CPUs of the servers without extra costs or performance loss. 

\subsection{Opportunity: CPU Peak Query Offloading} \label{sec:method}
According to the analysis in Section \ref{sec:concurrent}, NPUs/GPUs are prioritized to process queries to ensure performance quality. With the growth of query numbers, the maximum latency of processing in NPUs/GPUs increases, causing potential overtimes. Therefore, CPUs can be used to offload a small number of queries to mitigate timeout risks. The threshold to immigrate excessive queries to CPUs is the first hyperparameter to decide. The threshold $C_{NPU}^{max}$ must satisfy the following expressions:
\begin{gather} 
	t_{proc, NPU}^{C_{NPU}^{max}} \leq T    \label{eq:npu1} \\
	t_{proc, NPU}^{C_{NPU}^{max}+1} > T  \label{eq:npu2}
\end{gather}
Where $t_{proc, NPU}^{i}$ denotes the latency in NPUs with the system concurrency of $C_{NPU}^{max}$, and $T$ denotes the maximum allowed latency determined according to SLOs. The above expressions illustrate the scenario of having one more concurrent processed query causing overtimes. In that case, excessive queries can be immigrated to CPUs. Meanwhile, the upper boundary of query number process in CPUs must be set to avoid CPU timeouts. The upper boundary $C_{NPU}^{max}$ is determined using similar expressions as followings:
\begin{gather} 
	t_{proc, CPU}^{C_{CPU}^{max}} \leq T  \label{eq:cpu1}   \\
	t_{proc, CPU}^{C_{CPU}^{max}+1} > T  \label{eq:cpu2}
\end{gather} 

According to two deployment schemes mentioned in Equation \ref{eq:cost} and Equation \ref{eq:cost1}, we can analyze the theoretical saved costs of employing CPU-NPU collaboration method: if we deploy the system according to the maximum possible concurrent number, where both CPUs and NPUs have the maximum number of queries processed concurrently without any query waiting others to finish, there is no $t_{wait}$ for both CPUs and NPUs. Therefore, the system concurrency can be enlarged from $C_{NPU}$ to $C_{NPU} + C_{CPU}$, hence saving $\frac{C_{CPU}}{C_{CPU} + C_{NPU}} \times 100\%$ of deployment cost; if we use the average query numbers within a period of time, the system improves the average throughput up to $\frac{C_{CPU}}{C_{NPU}} \times 100\%$. Therefore, the maximum possible deployment cost decrement is $\frac{C_{CPU}}{C_{NPU}} \times 100\%$. 

Note that there is a scenario that CPUs cannot be used to offload excessive queries when processing a single query causes timeouts. That case has the following expression valid:  
\begin{equation}
	t_{proc, CPU}^{1} > T    \\
\end{equation}

\section{Peak Concurrent Queries Offloading}\label{sec:algorithm}
\begin{figure*}[!ht]
	\centerline{\includegraphics[width=0.9\textwidth]{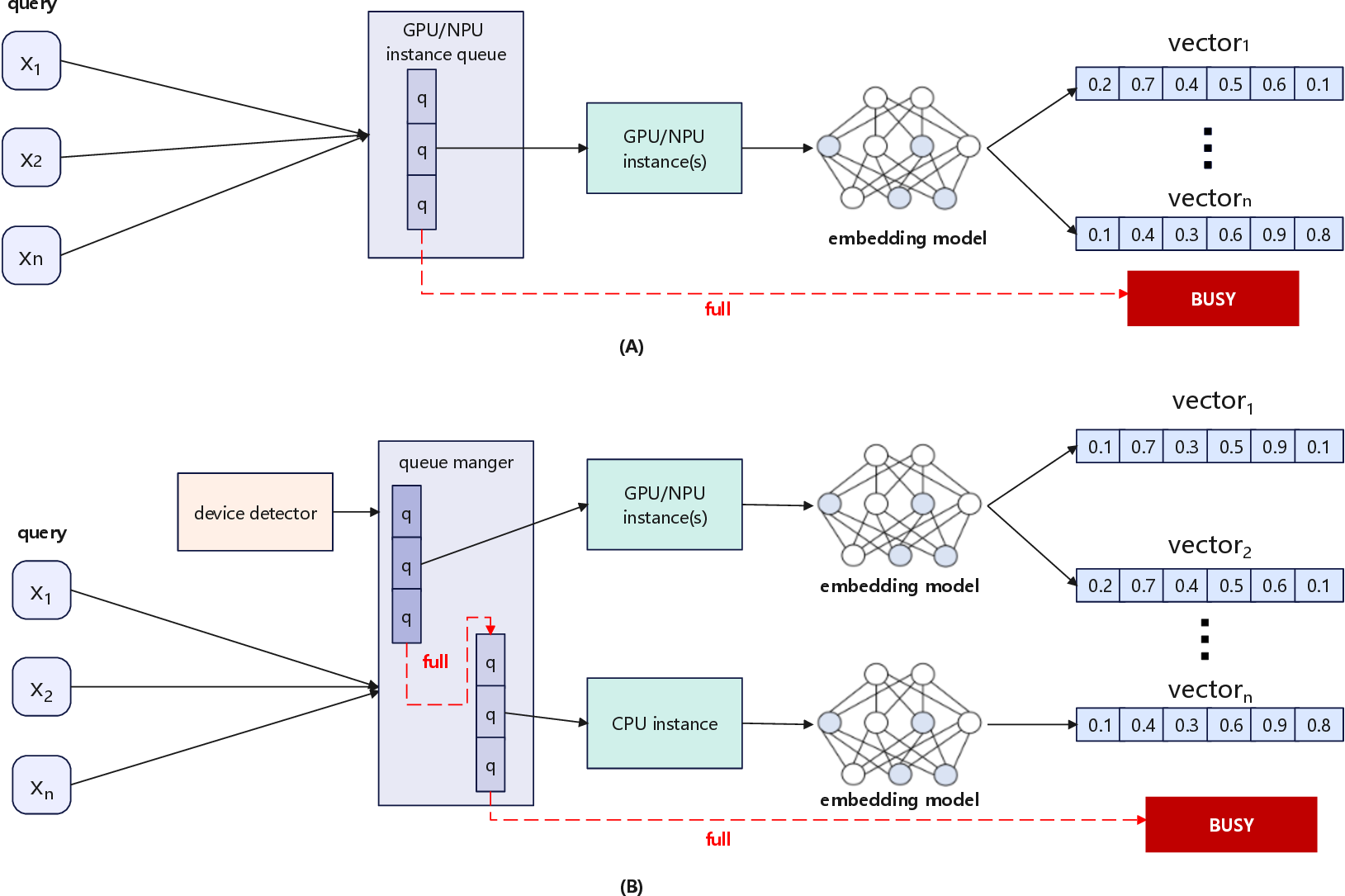}}
	\caption{WindVE System Compared to existing work. (A) shows a typical GPU/NPU-based vector embedding system workflow, and (B) shows the workflow of WindVE.                                                                                                                                                                                                                                                                                                                                                                                                                                                                                                                                                                     }
	\label{fig:system}
\end{figure*}
In Section \ref{sec:Motivation}, the importance of system maximum concurrency has been illustrated and the theoretical advantages of offloading peak queries to the CPU has been elucidated. Drawing upon these insights, the design methodology of WindVE is expounded upon in this section.

%\subsection{Design Methodology} \label{sec:method}

\subsection{System Design}
Figure \ref{fig:system} (A) illustrates how a conventional GPU/NPU-based vector embedding system works: GPU/NPU instances process queries retrieved from the GPU/NPU instance queue, generating vector representations as the output. To facilitate the offloading of peak queries to CPUs, WindVE employs a device detector in concert with a queue manager to streamline and complete the workflow, as illustrated in Figure \ref{fig:system} (B): upon service initialization, the device detector first identifies all available devices to instantiate embedding instances. Next, queues are established for each device to manage incoming queries, with the queue lengths set according to device thresholds ($C_{NPU}^{max}$ and $C_{CPU}^{max}$) decided in the previous subsection. 
When queries are received, they are allocated to the device queues based on the following dispatch policy: if only a single device is available, regardless of a CPU or an NPU/GPU, the queries are directed to this device unless its queue reaches the maximum capacity; with both CPUs and NPUs/GPUs operating in the service, the system prioritizes NPUs/GPUs to guarantee optimal performance. If the NPU/GPU queue reaches its limit, indicating that more queries could lead to timeouts, the system then routes them to the CPUs. If the CPU queue is also at full capacity, the service will decline excessive queries and respond with a `busy' status. Once queries are placed in the device queues, they are grouped into batches and processed by the corresponding instances. 
Each instance employs its own model copy to perform vector embedding. 

\subsection{Queue Manager}

\subsubsection{Overall Procedure}
The queue manager operates as a scheduler, responsible for dispatching incoming queries to suitable devices. To enhance the system’s maximum concurrency, the scheduling policy is structured as follows: priority is given to NPUs/GPUs for query processing to minimize latency. In the event of query bursts that surpass the established threshold, potentially leading to service timeouts, the excess queries are rerouted to the CPU instance, provided that the option for heterogeneous computing is activated. 
If the overflow of queries can also cause CPU timeouts, they are consequently rejected, and the service will respond with a ‘busy’ status. This policy effectively expands the system’s maximum capacity while maintaining the processing efficiency of NPUs/GPUs. The logic of the queue manager is detailed in Algorithm \ref{alg:query}.
\begin{algorithm}
	\small
	\caption{Query manager logic of CPU and NPU}
	\label{alg:query}
	\KwIn{$query_k, \forall k \in[1,K]$}
	\KwOut{The running instance for each query}
	\textbf{procedure} $QueryManager(query_k)$  \\
	\ForEach{$query_k$}
	{	
		\If{NPU queue is not full}
		{
			$npu\_queue\_length\ += 1$  \\	
			$npu\_queue\_push(query_k)$   \\
			$\Return \text{ `NPU'}$  \\
		}
		\Else
		{
			\If{heterogeneous computing is enabled}
			{
				\If{CPU queue is not full}
				{
					$cpu\_queue\_length\  += 1$  \\
					$cpu\_queue\_push(query_k)$   \\
					$\Return \text{ `CPU'}$  \\
				}
				\Else
				{
					$\Return \text{ `BUSY'}$  \\
				}
			}
			\Else
			{
				$\Return \text{ `BUSY'}$  \\
			}
		}
	}
\end{algorithm}

\subsubsection{Linear Regression Based Queue Depths Estimation} \label{sec:depth}
The pivotal question that remains is determining the appropriate depths for the queues. A straightforward method involves conducting stress tests in accordance with the methodologies outlined in Equations \ref{eq:npu1} and \ref{eq:npu2} for the NPU queue depth, as well as Equations \ref{eq:cpu1} and \ref{eq:cpu2} for the CPU queue depth. 
Specifically, the process increases the concurrency for standalone NPUs/GPUs and CPUs until the SLOs are no longer achievable. The maximum concurrency at which the SLOs are still met is noted as the potential queue depth. This procedure is then repeated with the CPU-NPU collaboration to refine the queue depths with greater precision. However, this process necessitates multiple rounds of profiling with incremental increases in concurrency. Particularly for devices capable of handling a large amount of concurrent queries, establishing the upper limit of concurrency can be a time-consuming endeavor. Moreover, the selection of the concurrency increment step is a delicate matter; too small a step may compromise efficiency, whereas too large a step risks overlooking the optimal maximum value.

To address this challenge, WindVE is equipped with a fast estimator designed to determine the upper limits of concurrency for both CPUs and NPUs/GPUs. The methodology is grounded in the observed linear relationship between the processing latency and the concurrency, as mentioned by SLSC \cite{slice} and Mooncake \cite{mooncake}. Therefore, the relationship between latency $t_{proc, d}^{C_{d}}$ and concurrency $C_d$, $d \in \{NPU, CPU\}$ can be expressed as follows:
\begin{equation} \label{eq:concurrency}
	t_{proc, d}^{C_{d}} = \alpha_{d} \cdot C_d + \beta_{d}, d \in \{NPU, CPU\}
\end{equation}
Here, $\alpha_{d}$ and $\beta_{d}$ present the fitting parameters with constraint of $\alpha_{d}, \beta_{d} \ge 0$. Consequently, leveraging this relationship, WindVE employs a method that involves conducting a limited number of profiling sessions. WindVE fits the correlation between processing latency and concurrency by applying Equation \ref{eq:concurrency}. Utilizing this linear regression, the system estimates the maximum concurrency that can be sustained while adhering to a specified SLO. The best queue depths can be fined-tuned based on the estimated values.

\subsubsection{Theoretical Effectiveness Analysis} \label{sec:analy}
Given the queue manager design methodology, the effectiveness of WindVE can be analyzed theoretically. Note that the latency is commonly split into three parts: the computing cost $t_{comp}$, the input/output moving cost $t_{io}$, and the model loading cost $t_{model}$, as illustrated in the equation below:
\begin{align} \label{eq:split}
	t_{proc} & = t_{comp} + t_{io} + t_{model}
\end{align}
Among the three contributing factors, $t_{comp}$ and $t_{io}$ exhibit a positive correlation with the concurrency $C_d$, whereas $t_{model}$ remains independent of $C_d$. In other words, the coefficient $\alpha_{d}$ in Equation \ref{eq:concurrency} is predominantly influenced by $t_{comp}$ and $t_{io}$ both of which are sensitive to the device’s computing power and memory bandwidth. Conversely, $\beta_{d}$ is mainly affected by $t_{model}$, which is solely affected by the memory bandwidth. %Equations below capture these features:
\begin{comment}
\begin{align}
	\alpha_{d} &= \frac{x_d}{T_d} + \frac{y_d}{M_d} \label{eq:alpha} \\
	\beta_{d} &= \frac{z_d}{M_d} \label{eq:beta}
\end{align}

Where $x_d, y_d, z_d$ are non-negative coefficients. %Therefore, we define a factor $\gamma_{d}$ to represent $\alpha_{d}$ to $\beta_{d}$ ratio:

\begin{align}
	\gamma_{d} &= \frac{\alpha_{d}}{\beta_{d}} \\
	& = \frac{x_d}{z_d} \cdot \frac{M_d}{T_d} + \frac{y_d}{z_d} \label{eq:ratio} 
\end{align}

From Equation \ref{eq:ratio}, a larger $\gamma_{d}$ means the device's memory bandwidth is less likely to become a bottleneck compared to its computing ability. 
\end{comment}

In CPU-NPU collaborative architectures, the following inequalities typically emerge as natural consequences of the CPU's constrained computational power and comparatively limited memory bandwidth:
\begin{align}
	\alpha_{CPU} &> \alpha_{NPU} \\
	\beta_{CPU} &> \beta_{NPU}
\end{align}

Therefore, we have the following derivations:
\begin{align}
	\beta_{CPU} &> \beta_{NPU} \\
	t - \beta_{NPU} &> t - \beta_{CPU} \\
	\alpha_{NPU} \cdot C_{NPU} &> \alpha_{CPU} \cdot C_{CPU} \\
	\frac{\alpha_{NPU}}{\alpha_{CPU}} &> \frac{C_{CPU}}{C_{NPU}} \label{eq:ineq}
\end{align}

As detailed in Section \ref{sec:method}, there is a direct correlation between the ratio $\frac{C_{CPU}}{C_{NPU}}$and the extent of extended concurrency. Inequality \ref{eq:ineq} delineates that the upper limit of this concurrency is governed by the ratio  $\frac{\alpha_{NPU}}{\alpha_{CPU}}$. Consequently, the narrower the performance disparity between NPUs/GPUs and CPUs, the greater the concurrency advantages that WindVE can leverage.

With further derivation from the ratio $\frac{\alpha_{NPU}}{\alpha_{CPU}}$, a set of new inequalities can be obtained:
\begin{gather}
	\frac{\alpha_{NPU}}{\alpha_{CPU}} > \frac{C_{CPU}}{C_{NPU}} \\
	\frac{\frac{\Delta t}{\alpha_{CPU}}}{\frac{\Delta t}{\alpha_{NPU}}} > \frac{C_{CPU}}{C_{NPU}} \\
	\frac{\Delta C_{CPU}}{\Delta C_{NPU}} > \frac{C_{CPU}}{C_{NPU}} \\
	\frac{C_{CPU} + \Delta C_{CPU}}{C_{NPU}+ \Delta C_{NPU}} > \frac{C_{CPU}}{C_{NPU}} \label{eq:loose}
\end{gather}
In this context, $\Delta t$ represents a positive increment in the time $t$; $\Delta C_{d}, C_{d}\in \{NPU, CPU\}$ signifies the increase in concurrency achieved when the SLO is relaxed from $t$ to $t+\Delta t$. Equation \ref{eq:loose} suggests that WindVE, when operating under a more lenient SLO, yields better performance benefits.

\subsection{Device Detector}
The device detector is used to detect available computing devices. The device queues are created according to the results of device detection and the settings of heterogeneous computing: if only one type of device is detected, only one queue will be created, and the heterogeneous computing option is forced to be disabled; if more than one type of devices are available but the heterogeneous computing option is not enabled, only NPUs/GPUs will establish a queue to ensure high performance; queues for CPUs and NPUs/GPUs are both initialized only if both types of devices are detected and the heterogeneous computing option is set. When devices are determined, corresponding worker numbers settings are loaded. WindVE recommends to have only one CPU instance per machine for lower latency. 
\begin{comment}
The device detector logic is shown in Algorithm \ref{alg:detect}.
\begin{algorithm}
	\small
	\caption{Device detector logic of CPU and NPU}
	\label{alg:detect}
	\KwIn{$NPU_i, CPU_j, \forall i \in[1,I], j \in [1,J]$}
	\KwOut{$device\_main,device\_auxiliary,worker\_num\_main$, \\
		$worker\_num\_auxiliary,heter\_enable$}
	\textbf{procedure} $DeviceDetect(NPU_i, CPU_j)$  \\
	$device\_main = \text{`none'}$ \\ 
	$device\_auxiliary = \text{`none'}$ \\
	$worker\_num\_main = 0$   \\
	$worker\_num\_auxiliary = 0$ \\	
	\If{npu is available}
	{	
		\If{heterogeneous computing is enabled}
		{
			/* heter\_enable=true */   \\
			$device\_main = \text{`npu'}$   \\
			$device\_auxiliary = \text{`cpu'}$ \\
			$worker\_num\_main = I$   \\
			$worker\_num\_auxiliary = J$ \\			
		}
		\Else
		{
			$device\_main = \text{`cpu'}$ \\
			$device\_auxiliary = \text{`none'}$ \\
			$worker\_num\_main = J$   \\
			$worker\_num\_auxiliary = 0$ \\				
		}
	}
	\Else
	{
		$device\_main = \text{`cpu'}$ \\
		$device\_auxiliary = \text{`none'}$ \\	
		$worker\_num\_main = J$   \\
		$worker\_num\_auxiliary = 0$ \\			
		$heter\_enable = false$	 \\
	}
	\Return $device\_main,device\_auxiliary,heter\_enable$  \\
\end{algorithm}
\end{comment}
\subsection{CPU Affinity and Numa Settings for ARM CPUs}
A phenomenon has been noticed when WindVE was practically implemented and tested: CPUs in ARM architecture showed obvious performance improvement with core affinity assignment. CPU affinity refers to binding a process to a specific CPU core or a set of cores, to avoid the unnecessary communication costs caused by frequent core switch in a single process when CPU affinity is not set. 

Another phenomenon was that assigning cores with large indices performs better than with small indices. This is because the service framework runs on cores with small indices by default. 

Meanwhile, CPU cores are split into several groups called `numas', each of which have cores sharing an individual block of memory. Cores accessing the memory block associated to their numa has higher speed than accessing memory blocks belong to other numas. Therefore, avoiding cross-numa CPU affinity setting is also crucial in improving CPU performance. 

According to what has been observed in real practice, an empirical suggestion has been conducted for better performance: CPUs in ARM architecture should assign CPU affinity with core indices in reversed order and without numa crossing if possible.
\section{Experiments}\label{sec:experiment}
In this section, a series of experiments for overall performance and the system scalability have been delivered on the CPU-NPU collaborative system. Experiment settings are introduced in Section \ref{sec:settings}. 
These settings serve as defaults in experiments for overall performance demonstrated in Section \ref{sec:stresstest} and \ref{sec:scalability} for system scalability if not mentioned specifically.

\subsection{Experiment Settings} \label{sec:settings}

\subsubsection{Testbed}
To validate the CPU-NPU collaborative system, experiments have been run in CPUs with both x86 and ARM architectures, and in both NPUs and GPUs. Two type of machines have been employed in experiments: one is a GPU machine with Tesla V100 GPUs \cite{tesla-100} and Intel Xeon E5-2680 CPUs \cite{intel-xeon-e5-2680} with x86 architecture and 48 cores in total; another is Atlas 800 Inference Server (Model: 3000) \cite{atlas3000} with Atlas 300I DUO NPUs \cite{atlas300iduo} and Kunpeng 920 CPUs \cite{kunpeng920} with ARM architecture and 128 cores in total.

\subsubsection{Embedding Models} 
The mainly embedding model used in the experiments is bge-large-zh-v1.5 model \cite{bge}. Bge-large-zh-v1.5 is an embedding model to convert input queries to a output vector with 1024 \textit{fp32} elements, with around 326 million parameters. 
Besides, jina model (\cite{jina}, 570M parameters and 8192 output length) is also adopted as supplementary.

\subsubsection{Workloads } 
The length rather than the content of input queries matters for vector embedding service. 
In all experiments except for the scalability with query length, the default query length is 75 tokens, which is a typical setting for text segmentation in RAG systems. 
Input queries are sent concurrently and organized in batches. 
A new batch of queries will be sent only after the responses of previous batches have been received.

\subsubsection{Baselines}
The embedding framework used in experiments is FlagEmbedding \cite{site:flagembed}, which is based on PyTorch. 
FlagEmbedding is an embedding framework using BGE (BAAI General Embedding), a toolkit focused on retrieval-augmented LLMs with a set of projects in inference, fine-tuning, evaluation, datasets, tutorials, and research. 
As for jina embedding model, it can be executed directly by the \textit{transformer} module of PyTorch.

\subsubsection{Implementations}
To utilize the CPU processing without slowing down the tasks in NPUs, we reserved 1/4 of the total cores for Arm-based Kunpeng CPUs with low indexes for NPU processing. Additionally, we ensured the NPU’s maximum concurrency when utilizing CPU offloading to avoid sacrificing NPU processing.

\subsubsection{Experiment Procedures}
The overall performance evaluation of WindVE is presented, juxtaposed against the baseline under the metric of the maximum concurrency given two typical SLOs in real-world business applications: e2e latency not exceeding 1s and 2s, respectively. 
Next, the queue depth estimator is evaluated individually, contrasted with the approach of utilizing stress tests to ascertain queue depths. Finally, the impact of varying input query lengths and the number of employed CPU cores is explored.

\textbf{\begin{table}[h]
		\begin{tabular}{l|cc|cc}
			\hline
			NPU/GPU device              & \multicolumn{2}{c|}{Tesla V100}         & \multicolumn{2}{c}{Atlas 300I DUO} \\ \hline
			CPU device                  & \multicolumn{2}{c|}{Intel Xeon E5 2690} & \multicolumn{2}{c}{Kunpeng 920}    \\ \hline
			time limits (s)             & \multicolumn{1}{c|}{1}        & 2       & \multicolumn{1}{c|}{1}      & 2     \\ \hline
			FlagEmbedding conc.   & \multicolumn{1}{c|}{44}       & 96      & \multicolumn{1}{c|}{84}     & 172   \\ \hline
			WindVE conc. & \multicolumn{1}{c|}{44 + 8}        & 96 + 22     & \multicolumn{1}{c|}{84 + 1}      & 172 + 8    \\ \hline
			conc. improvement     & \multicolumn{1}{c|}{18.2\%}   & 22.3\%  & \multicolumn{1}{c|}{1.2\%}  & 4.7\% \\ \hline	
		\end{tabular}
		\caption{Overall performance of WindVE against FlagEmbedding ob bge model  under 1s and 2s limit.}
		\label{tab:estimate}
\end{table}}
\vspace{-0.4cm}
\textbf{\begin{table}[h]
		\begin{tabular}{l|cc|cc}
			\hline
			NPU/GPU device              & \multicolumn{2}{c|}{Tesla V100}         & \multicolumn{2}{c}{Atlas 300I DUO} \\ \hline
			CPU device                  & \multicolumn{2}{c|}{Intel Xeon E5 2690} & \multicolumn{2}{c}{Kunpeng 920}    \\ \hline
			time limits (s)             & \multicolumn{1}{c|}{1}        & 2       & \multicolumn{1}{c|}{1}      & 2     \\ \hline
			PyTorch conc.   & \multicolumn{1}{c|}{48}       & 112      & \multicolumn{1}{c|}{128}     & 256   \\ \hline
			WindVE conc. & \multicolumn{1}{c|}{48 + 11}        & 112 + 30     & \multicolumn{1}{c|}{128 + 6 }      & 256 + 20    \\ \hline
			conc. improvement     & \multicolumn{1}{c|}{22.9\%}   & 26.7\%  & \multicolumn{1}{c|}{4.6\%}  & 7.8\% \\ \hline		
		\end{tabular}
		\caption{Overall performance of WindVE against PyTorch on jina model under 1s and 2s limit.}
		\label{tab:estimate-jina}
\end{table}}

\begin{figure*}[htp]
	\centerline{\includegraphics[width=0.9\textwidth]{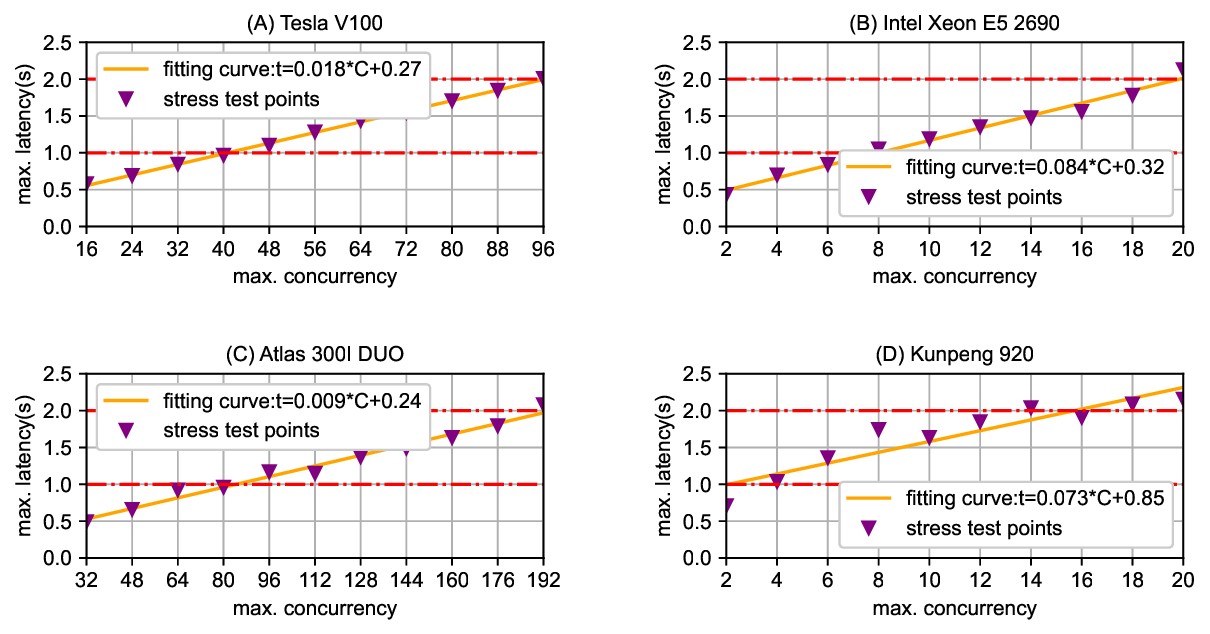}}
	\caption{Fitting curves and stress tests on (A) Tesla V100 GPU, (B) two Intel Xeon E5 2690 CPUs, (C) one Atlas 300I DUO NPU, (D) and two Kunpeng 920 CPUs. }
	\label{fig:stresstest}
\end{figure*}

\begin{table*}[h]
	\begin{tabular}{cc|cccc}
		\hline
		\multicolumn{2}{c|}{device queue depth}                                                         & Tesla V100 & Intel Xeon E5 & Atlas 300I DUO & Kunpeng 920 \\ \hline
		\multicolumn{1}{c|}{\multirow{3}{*}{1s limit}} & linear regression   & 40         & 8             & 84             & 2           \\ \cline{2-6} 
		\multicolumn{1}{c|}{}                                                 & stress test & 40         & 6             & 80             & 2           \\ \cline{2-6} 
		\multicolumn{1}{c|}{}                                                 & fine-tuned      & 44         & 8             & 84             & 2           \\ \hline
		\multicolumn{1}{c|}{\multirow{3}{*}{2s limit}} & linear regression   & 96         & 20            & 195            & 15          \\ \cline{2-6} 
		\multicolumn{1}{c|}{}                                                 & stress test & 88         & 18            & 176            & 12          \\ \cline{2-6} 
		\multicolumn{1}{c|}{}                                                 & fine-tuned   & 96         & 22            & 172            & 8           \\ \hline
	\end{tabular}
	\caption{Queue Depth Prediction through linear regression and through stress tests.}
	\label{tab:queuedepth}
\end{table*}

\begin{figure*}[h!]
	\centerline{\includegraphics[width=0.9\linewidth]{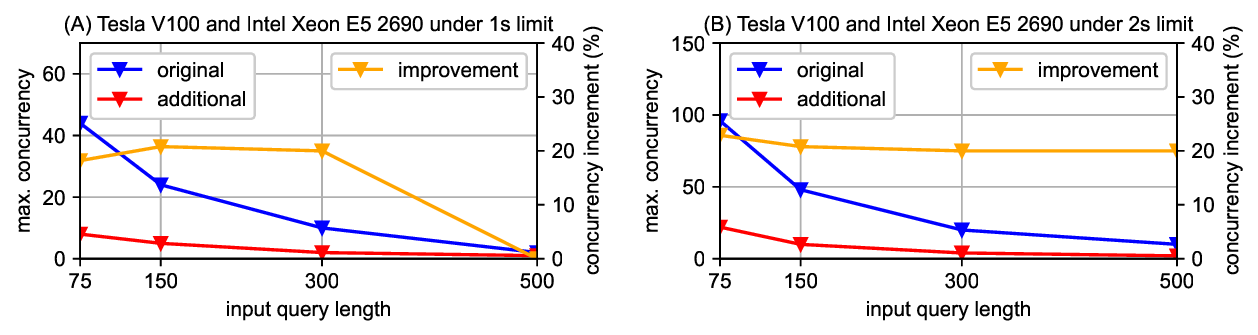}}
	\caption{Scalability experiment with varied input query length in Tesla V100 and Intel Xeon E5 2690 under 1s and 2s limit.}
	\label{fig:length}
\end{figure*}
\begin{figure*}[h!]
	\centerline{\includegraphics[width=0.9\linewidth]{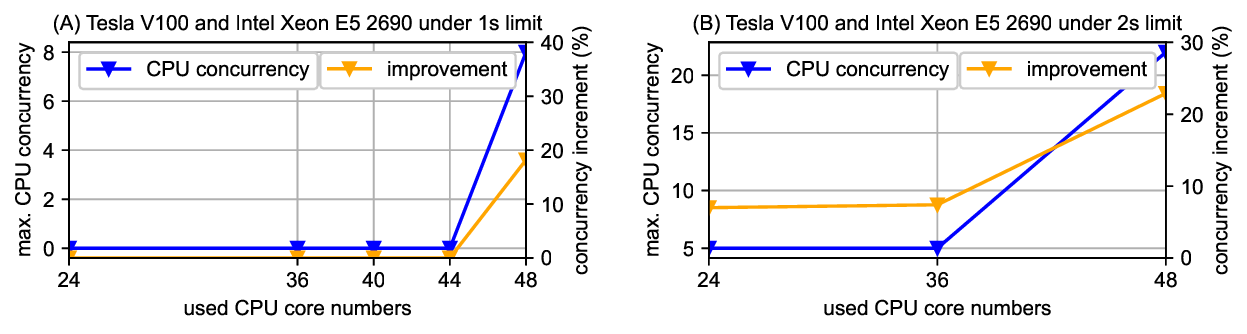}}
	\caption{Scalability experiment with varied CPU core numbers in Tesla V100 and Intel Xeon E5 2690 under 1s and 2s limit.}
	\label{fig:cores}
\end{figure*}
\subsection{Overall Performance} \label{sec:stresstest}

Initially, we employ the queue depth estimator detailed in Section \ref{sec:depth} to estimate the possible maximum concurrency for NPUs/GPUs and CPUs.
Next, the queue depth is fine-tuned according to the estimated value with CPUs and NPUs/GPUs running collaboratively. 
The overall performance of WindVE against counterparts on \textit{bge} and \textit{jina} models are shown in Table \ref{tab:estimate} and Table \ref{tab:estimate-jina}, respectively. 

\Paragraph{Comparison on bge model}.
From the results, the best concurrency improvement of 22.3\% is given by the collaboration of one Tesla V100 GPU and two Intel Xeon E5 2690 with a less than 0.1 ms overhead. Besides, by applying CPU offloading, the average CPU utilization rises from 8\% to 80\%.
Based on the analysis given in \ref{sec:method}, this improvement can reduce 18.6\% deployment cost when using maximum concurrent query numbers to deploy, and save up to 22.3\% deployment cost when using average query numbers. 
% on average case, why it is 22.3%

\Paragraph{Comparison on jina model}.
Apparently, both CPU and NPU/GPU exhibit more concurrency on jina than bge, e.g., the concurrency of Kunpeng 920 is 20 when the time limit is 2s, much higher than the value (8) on bge. The overhead stays less than 0.1ms, and the average CPU utilization rises from 9\% to 81\%.
The concurrency improvement reaches 26.7\% with Intel Xeon E5 2690, compared to the baseline that uses Tesla V100 only.
Thus, the deployment cost of jina model can be reduced by 21.1\% and 26.7\% on maximum and average cases, respectively. 

Three notable phenomena emerge from Table \ref{tab:estimate}: 
\begin{itemize}
\item the concurrency improvement within a 2s limit surpasses that of a 1s limit across both device combinations;
\item the combination of Tesla V100 GPU and two Intel Xeon E5 2690 exhibits a greater concurrency improvement compared to the Altas 300I DUO and Kunpeng 920 combination under both time constraints;
\item on an embedding model with faster speed of inference, the concurrency improvement is more prominent on both CPU and GPU/NPU, as well as 1s/2s time limit.
\end{itemize}
Section \ref{sec:analy} has already furnished the theoretical underpinnings for the existence of these phenomena, and this section delivers experimental evidence to validate the analysis.

Figure \ref{fig:stresstest} further illustrates the fitting curves for all four devices involved in the experimental setup. 
A comparison between Figure \ref{fig:stresstest} (A) and (B), as well as (C) versus (D), reveals that the fitting coefficient $\beta_{CPU}$ (0.32 for the Intel Xeon E5 2690 in Figure \ref{fig:stresstest} (B) and 0.85 for the Kunpeng in Figure \ref{fig:stresstest} (D)) consistently exceeds $\beta_{NPU}$ (0.27 for the Tesla V100 in Figure \ref{fig:stresstest} (A) and 0.24 for the Atlas 300I DUO in Figure \ref{fig:stresstest} (C)). 
Consequently, as the time constraint relaxes from 1 second to 2 seconds, the concurrency improvement increases, rising from 18.2\% to 22.3\% for the Tesla V100 and Intel Xeon E5 2690, and from 1.2\% to 4.7\% for the Atlas 300I DUO and Kunpeng 920. 
This empirical data validates the prediction that a looser SLO facilitates a more significant enhancement in concurrency.

Furthermore, the ratio of $\frac{\alpha_{NPU}}{\alpha_{CPU}}$ for the Tesla V100 and Intel Xeon E5 2690 is greater than that of the Atlas 300I DUO and Kunpeng 920 (0.21 versus 0.12). This suggests that the performance disparity between the Tesla V100 and Intel Xeon E5 2690 is less pronounced than that between the Atlas 300I DUO and Kunpeng 920. Consequently, this results in a more substantial concurrency enhancement for the Tesla V100 and Intel Xeon E5 2690 compared to the Atlas 300I DUO and Kunpeng 920. This experimental results prove the validation of the prediction that a smaller performance gap between NPUs/GPUs and CPUs allows better concurrency improvement.

\subsection{Queue Depth Estimator Evaluation}
An individual assessment of the queue depth estimator has been conducted, and the results are presented in Table \ref{tab:queuedepth}, showcasing the queue depth predictions produced from linear regression, stress tests with the increment step of 8, and the final fine-tuned queue depth achieved during CPU-NPU collaboration. The results indicate that the predictions yielded by linear regression demonstrate either superior or comparable efficacy to those obtained through stress tests, with the exception of the cases involving the Atlas 300I DUO and Kunpeng 920 under the 2-second limit. The comparatively better performance of the linear regression based prediction is attributed to an overly large increment step in the stress tests, which resulted in missing the peak of maximum concurrency. Nevertheless, Atlas 300I DUO and Kunpeng 920 generate a larger number of outliers when mapping the relationship between concurrency and end-to-end latency. This results in a less accurate prediction of queue depths when compared to the results of stress testing.

\subsection{Scalability Experiments} \label{sec:scalability}
Experiments for scalability have also been delivered with varied input query length and CPU core numbers in Tesla V100 and Intel Xeon E5 2690. 
Figure \ref{fig:length}, %Table \ref{tab:multinpu} 
and Figure \ref{fig:cores} give the results with varied query length and used cpu core numbers,  with NPU/GPU number fixed.

\Paragraph{Scalability with query length}.
In Figure \ref{fig:length}, the ``original'' curve denotes the concurrency on NPU only, while the ``additional'' curve denotes the additional concurrency brought by CPU.
Obviously, longer queries degrade the concurrency both in CPUs and NPUs/GPUs. 
With the time limit set to 1s, the additional concurrency decreases to 0 when the input query length reaches 500, as the CPU embedding can no longer meet SLOs.
In contrast, with the time limit set to 2s, our method still has 2 additional concurrency and 20\% improvement even when the query length is 500.  
This is because longer sequences amplifies the performance difference between NPUs/GPUs and CPUs. 
On the one hand, with smaller SLO, the degradation is more fast and the concurrency is lower.
On the other hand, the drop of concurrency improvement is not particularly obvious when the performance gap of different devices is not prominent (e.g., Tesla V100 and Intel Xeon E5 2690). 

\Paragraph{Scalability with CPU cores}.
Figure \ref{fig:cores} shows that fewer CPU core numbers worsen the CPU concurrency on Intel Xeon E5 2690. 
With 1s limit, using less than 44 CPU cores does not bring any performance benefit, because the loss of computing ability leads to a surge of CPU latency.
This boundary can be reduced to 36 given 2s limit, as these CPU cores are enough to finish the embedding within 2s.
In practical, with 2 CPUs with 128 cores (usually organized as 4 numas), we can utilize at most 96 cores (corresponding to the latter 3 numas) because our main program as well as the system programs runs on the first numa.
However, even with more CPUs equipped in the server, the concurrency can not be improved continuously after a border, due to the bottleneck of host memory bandwidth (only tens of Gigabytes per second).
% also impacted by the batch size and query length

As a summary, having longer input queries, and less CPU cores all deteriorates the effects the CPUs; 
In actual practice, WindVE is more likely to achieve higher benefit with short queries, with as many cores as possible, and with small CPU-to-NPU performance gap. 
%Experiments with changing query lengths, CPU core nums, NPU device nums, models, etc

\section{Conclusions}\label{sec:conclusion}
In this paper, we identifies the relationship between embedding service costs, throughput, and concurrency. To address deployment delays caused by average processing times, we developed a queue manager that offloads excess queries to idle CPUs, boosting system concurrency without slowing processing. Theoretical guidance helps calibrate its key parameter for maximum concurrency. This manager forms the core of WindVE, a system designed to increase throughput via concurrency, unlike others focusing on latency. Mathematical expressions allow performance estimation and hyperparameter tuning. Experiments show WindVE boosts concurrency by up to 22.3\% using one GPU and two CPUs. Scalability tests indicate WindVE is most effective with short queries, ample CPU cores, and minimal CPU-GPU performance gaps.

\bibliographystyle{IEEEtran}
\bibliography{sample-base} 

\end{document}